# Tangible Music Interfaces Using Passive Magnetic Tags

**Joseph A. Paradiso, Kai-yuh Hsiao, and Ari Benbasat**
Responsive Environments Group
MIT Media Laboratory
Cambridge, MA  02139
+1 617 391 6107
joep@media.mit.edu

**ABSTRACT**
The technologies behind passive resonant magnetically-coupled tags are introduced and their application as a musical controller is illustrated for solo or group performances, interactive installations, and music toys.

**INTRODUCTION**
Although noncontact gesture sensors [1], generally based around capacitive sensing [2] or computer vision and optical approaches [3], make very expressive musical controllers, they suffer from a lack of a tactile interface (leading to a deficit in precise, virtuosic input) and often the inability to reliably identify and distinguish different objects or body parts (limiting variation in control).  The combination of free-gesture sensing with a tangible reference has the potential of producing a very expressive electronic musical interface that also encompasses a degree of tactile precision and versatility. A generic example of this kind of combination is our digital baton [3], a handheld device that sported a set of accelerometers and a precise optical tracker for motion and position measurement together with several continuous pressure sensors for deliberate finger input.  Although the handheld electronics were relatively compact, this was an active device of considerable complexity.  Another approach of this sort is represented in the "Augmented Groove" project [4] by ATR and the University of Washington, where a computer vision system recognized a set of objects based on the appearance of a printed label and, when detected, inferred their position and orientation.  Each object  was used to produce or modify a particular set of musical sounds and effects, which were modulated as the object was manipulated.  The objects (simple vinyl LP records in this case) were passive and extremely simple, while the large variety of different targets identified by the system led to a wide range of control parameters and produced a natural collaborative environment for multiplayer interaction.

As these examples exploit optical tracking, they require a clean line of sight, and are subject to occlusion by body parts or other objects in the vision field.  One approach that doesn't suffer from this problem is magnetic tracking, as relatively low-frequency magnetic fields pass easily through the body and  other non-ferrous entities.   The computer graphics industry has perfected such techniques for motion capture applications; these trackers, while quite accurate and sensitive across many meters of range, require active pickups, which include potentially complex hardware and often a wired connection to a base station or master "beltpack".  On the other hand, passive RFID tags are very simple transponders that are powered by magnetic, electrostatic, or RF  energy provided by a "reader" basestation antenna.   These tag packages tend to be extremely compact and inexpensive.  By placing such a "tag" on each object that one uses as a controller, one attains the potential of identifying and tracking them without needing a clear line-of-sight.  Going further, if these tags respond to a local physical parameter (e.g., pressure, finger position, etc.), the objects can be made to exhibit additional degrees of tactile expression.

Commercial chip-based RFID systems, while capable of addressing virtually limitless ID space, generally tend to respond too slowly for realtime musical control, especially when multiple tags are present, where anticollision protocols can lead to considerable delay.   Simple magnetically-coupled resonant tags (i.e., only an electrical or mechanical resonance without a CMOS chip) don't necessarily have this problem; if each tag has a different resonant frequency, they can all be read quickly  without interference.  On the other hand, the space of available ID's is limited (determined by the Q of the resonance and the amount of frequency sweep), but still ample for distinguishing several dozen objects, which is adequate for an interesting musical controller.  Since the coupling of the tags with the local magnetic field is dependant on orientation, objects with three orthogonal tags are able to determine orientation (relative to the local field vector) independently from range (orientation can also be decoupled from range by using multiple reader coils).  In addition, resonances can be made to be parametric with a local mechanical parameter such as pressure (which can detune the coil or change the capacitance of a LC resonator, for example), enabling them to also function as tactile sensors.

Although resonant tags are most commonly used in Electronic Article Surveillance (EAS) applications as shoplifting tags, they have made recent, limited inroads into entertainment, showing up in board games [5] and in the mallets of Don Buchla's "Marimba Lumina" MIDI controller [6]. These are both close-range interactions, however, with all gesture being sensed within an inch or so of an interactive plane and precise position tracked across the surface - descendants of the familiar Wacom whiteboard [7], which similarly used magnetic tags to track and identify a set of coded pens. Our research in using tags as a musical controller conversely looked at measuring the tags across a larger volume (e.g., a 1-foot cube) and responding to orientation, proximity, and local parameters (e.g., pressure).

### TAG READER TECHNOLOGIES

We have investigated two different types of tag readers for this research, as detailed in [8]. The first, diagrammed in Fig. 1, is a "pulse-induction ringdown" reader, where we send out a brief magnetic pulse at the resonant frequency of a tag and then listen for a ringdown response. This mode of detection is very sensitive to high-Q resonators, such as the magnetostrictor tags used in Sensormatic's popular UltraMax EAS system. For maximum efficiency, the transmit coil is made resonant, tuned dynamically to different tags by adding discrete capacitors through a triac-switched, binary-weighted ladder.

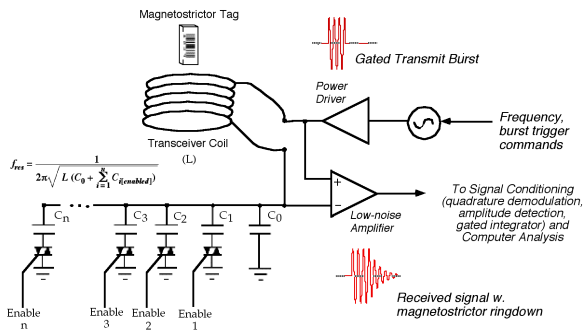

**Figure 1: Basic design of a pulse-induction ringdown reader**

Although our prototype ringdown reader functioned well and exhibited high sensitivity that promised an extensive range of detection, this technique was intrinsically slow, requiring on the order of 5-10 milliseconds to read each tag (including time to tune the reader, pulse the tag, integrate and digitize the received signals, etc.). For multiple tags, this response speed can become problematic. Although one could excite several tags simultaneously with a broadband burst (such as used in NMR techniques), a nonresonant transmit coil leads to difficult driver inefficiencies here.

We have thus moved to a different tag reader system based around detecting the dynamic loading that the tags present to a reader coil driven by an exponential FM chirp that spans a decade of frequency. This system has evolved over the last few years; it was introduced in [9] and described in [10] and [11]. The reader's block diagram is shown in Fig. 2 - this design likewise has a lineage that descends from shoplifting systems. We currently detect the dynamic loading of the tags on the reader coil by an inductive bridge, subsequently shaping and detecting the resultant signal through a bank of highpass filters. An embedded microcontroller isolates the peaks created by nearby tags and sends their parameters across a serial line to a host computer at the 30 Hz chirp rate. Our current system sweeps from roughly 40 kHz to 400 kHz, and can detect both magnetostrictors (resonating below 100 kHz) and LC (inductor-capacitor) tags. Although we currently use an analog oscillator to sweep, the microcontroller can periodically calibrate any frequency drift by counting oscillator cycles with the control voltage held stable.

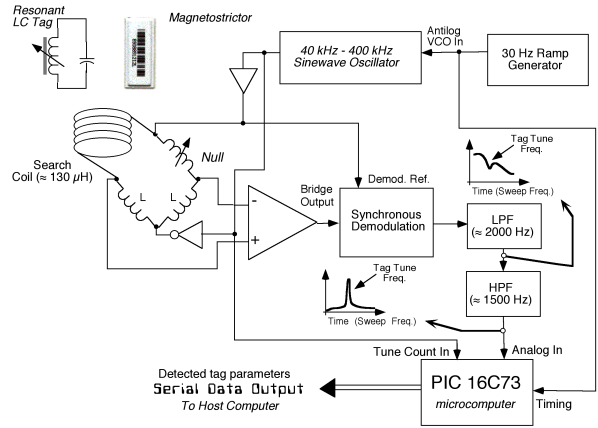

**Figure 2: Swept-frequency tag reader**

Figure 3 shows a picture of our current tag reader board and reader coil - the peaks visible on the scope result from the tags that are resting inside the coil.

### THE TAGGED OBJECTS

Our current research and demonstration system uses an array of 16 tagged objects, as shown in Figure 4. Two (the cube and eyeball) have three embedded tags to determine full orientation together with proximity. One (the Pez) can also detect the position of its head as it's pulled (a ferrite bead is attached to the dispenser rod, which shifts the tune of a coil wound around the Pez's body). The "ghosts" have a plastic band attached so they can be worn as rings, enabling each finger to provide independent control.

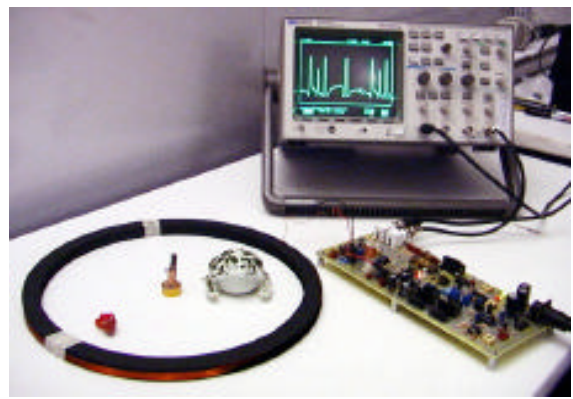

**Figure 3: Swept reader board, coil, tags, and scope response**

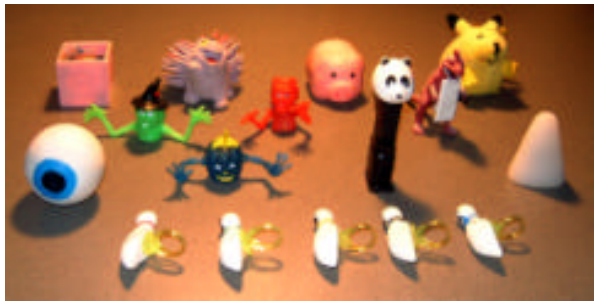

**Figure 4: Current ensemble of tagged objects**

All objects are very simple, made from nonferrous, nonconductive plastic or resin, either cast by us or purchased as "trinkets" at a local toy store. Although any of these objects can be picked up and moved freely, the design of each particular object lends itself to particular applications; some are structured to be able to sit atop the reader's table, launching background sounds and textures, while others can be rolled around, dynamically adding and modifying audio and graphic effects.

**MUSICAL APPLICATIONS**

A simple musical environment has been assembled to demonstrate the possibilities of this interface. This installation, appropriately named *Musical Trinkets*, was inspired by the early performances of the composer/producer John Zorn, who would give extended concerts of improvised music by "playing" several acoustic sound-producing objects (e.g., bird calls, shakers, percussive things) strewn across a table. In *Musical Trinkets*, we do this digitally, where each tagged object acquires a set of complex musical properties when in proximity to the reader. An attached PC laptop maps the dynamic tag data into a MIDI stream to drive a set of music synthesizers and produces accompanying graphics.

In particular, the goblins launch background droning chords (with amplitude proportional to their distance from the reader) and set the notes that are played by the velocity-sensitive rings, which each fire when brought close to the reader and into alignment with the solenoidal magnetic field. The Pengachu plays a sequence of textural notes; their pitch rising with proximity. The cube launches a low, droning tone, with pitch smoothly bending as a function of orientation (sounding much like Jon Hassel's processed trumpet). The Pez introduces a choral voice, firing another sound (e.g., a synthesized brass) when pulled, with filter cutoff depending on the extent of the pull. The other objects are modifiers, which affect the active voices. The porcupine is a pitch shifter (bending pitch continuously down with proximity), the pig produces vibrato, and the eyeball introduces audio effects (each of the three tags in the eyeball controls a different parameter in a complex patch running on an attached Lexicon LXP100). Although tags with adjacent frequencies can have some limited interactions (e.g., one tag can somewhat suppress the sensitivity of the other), all can be used together.

A couple of graphical environments have also been written for this system. One is a very simple GL mapping, with simple geometric shapes and actions driven directly by the tag data. Another is a much more complicated behavior-based system called "Music Creatures"[12] that produces cloud-like streamers driven from the MIDI data. The graphics are projected onto a frosted rear-screen inside the reader coil (see Fig. 5), hence provide direct feedback to the user manipulating the objects above. The "triangle" tag switches between the two graphics systems. More details on the graphical and musical mappings are given in [8]. This installation has been publicly exhibited twice thusfar, first at SIGGRAPH 2000 (Fig. 6) and several months later to very large crowds at SMAU in Milan (Fig. 7).

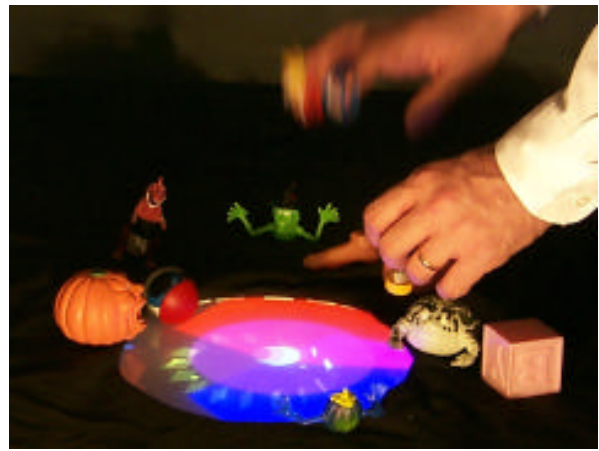

**Figure 5: Musical Trinkets in action**

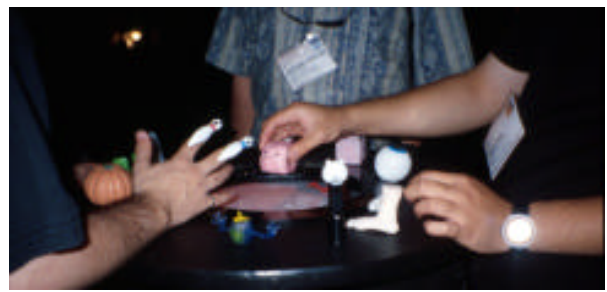

**Figure 6: Musical Trinkets at SIGGRAPH 2000**

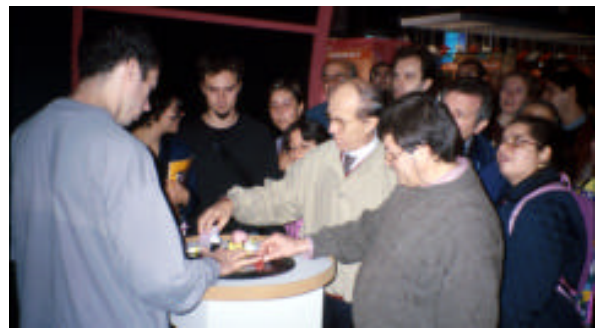

**Figure 7: Musical Trinkets at SMAU in Milan**

The reaction to this installation has generally been very positive - people enjoy sculpting sounds and graphics with our toolkit of objects. The mapping, although simple, isn't entirely intuitive, however, and this installation generally requires an attendant to get participants started.

After living with our *Musical Trinkets* environment for several months, it has become obvious to us this mapping stays at too basic a level - it hints at possibilities for virtuosic performance, but despite the variety of objects available, it often stays in too simple a sonic space. We are now looking at other mapping strategies to lend greater depth, turning this from an engaging toy into a viable performance interface. These include looking at complex context shifts as more tags are introduced (instead of dedicating particular effects to particular tags), using the finger controllers for multiparameter effects instead of just note launches, and the ability to dynamically record and playback sequences and textures, associating them with particular objects. As this interface is very well suited to group improvisation (witness Figs. 6 and 7), hence we'll be exploring mappings that encourage such interaction.

### HARDWARE EVOLUTION

We are also exploring modifications to our hardware to extend its performance. One direction is to increase the reader's sensitive range, enabling interaction across person-sized volumes (creating, for instance, a very interesting musical environment for a juggler). Along these lines, we have recently built simple active tags that sense across volumes of several meters (yielding battery life of several days and not requiring tethers) [13].

Another direction that we're pursuing is the application of tag techniques to precise position and orientation sensing. With our existing single-coil solenoid, the magnetic flux loops around the coil to return on its backside, producing a correlation between position and orientation (Fig. 8a). We have made an arrangement of 3 perpendicular Hemholtz coils to avoid this problem, as depicted in Fig. 8b and shown in Fig. 9 as a working prototype. Here, we drive opposing coils simultaneously, pulling the magnetic flux directly across, making a more uniform field distribution. This creates a controller, where tagged objects (or a hand with tag rings) can be inserted into a wireframe cube, where they are identified and precisely tracked. We are beginning to explore musical applications of this system, both as a "tangible" controller on its own and as a "left-hand" controller augmenting performance on another instrument

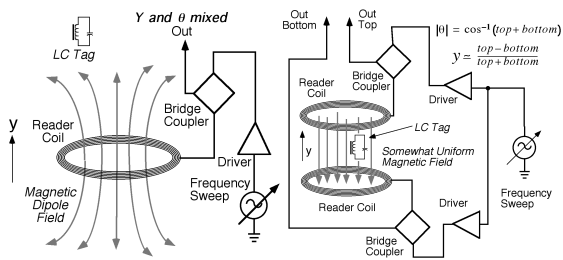

**Figure 8: a) Existing solenoidal system (left) and b) new Hemholtz tracker volume (right)**

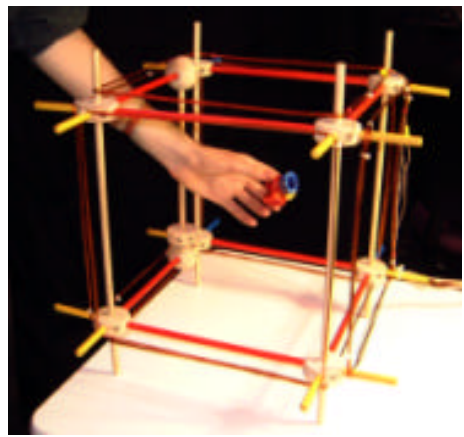

**Figure 9: Working prototype of precision tag tracker**

### ACKNOWLEDGMENTS
We thank our many colleagues who have been involved in this project, especially Dan Stiehl, Marc Downie, and Josh Lifton. We appreciate the support of TTT Consortium and other sponsors of the MIT Media Lab.

### REFERENCES

1. Paradiso, J. "Electronic Music Interfaces: New Ways to Play," *IEEE Spectrum*, 34(12), 1997, pp. 18-30.

2. Paradiso, J., Gershenfeld, N. "Musical Applications of Electric Field Sensing", *Computer Music Journal*, 21(3), 1997, pp. 69-89.

3. Paradiso, J., Sparacino, F., "Optical tracking for music and dance performance", in *Optical 3-D Measurement Techniques IV*, A. Gruen, H. Kahmen Eds. Herbert Wichmann Verlag, Heidelberg Germany, 1997, pp. 11-18.

4. Poupyrev, I., "Augmented Groove: Collaborative Jamming in Augmented Reality," in *SIGGRAPH 2000 Conference Abstracts and Applications,* ACM Press, NY, p. 77.

5. http://www.zowiepower.com/about/smart2.html#technoloy

6. O, Larry the, "Marimba Lumina: This is not your mother's MIDI controller" Electronic Musician, 16(6), June 2000.

7. Murakami, A., et al, "Position Detector," US Patent No. 5,466,896, Nov. 14, 1995.

8. Hsiao, K., "Fast multi-axis tracking of magnetically resonant passive tags: methods and applications," MS Thesis, MIT Department of Electrical Engineering and MIT Media Laboratory, February 2000.

9. Paradiso, J. and Hsiao, K., "Swept-Frequency, Magnetically-Coupled Resonant Tags for Realtime, Continuous, Multiparameter Control," *CHI99 Extended Abstracts*, ACM Press, NY, 1999, pp. 212-213.

10. Hsiao, K. and Paradiso, J. "A New Continuous Multimodal Musical Controller Using Wireless Magnetic Tags," *Proc. of the 1999 International Computer Music Conference*, October 1999, pp. 24-27.

11. Paradiso, J., *et al*, " Sensor Systems for Interactive Surfaces," *IBM Systems Journal*, Volume 39, Nos. 3 & 4, October 2000, pp. 892-914.

12. Downie, M., "Behavior, Animation and Music: The Music and Movement of Synthetic Characters," MS Thesis, MIT Media Laboratory, January 2001.

13. http://acg.media.mit.edu/projects/atmosphere/